\documentclass[MSNbibl,nameyear,dvips]{arxstspdf}
\usepackage{flushend}
\usepackage{stfloats}
\usepackage{graphicx}

\volume{27}
\issue{1}
\pubyear{2012}
\firstpage{24}
\lastpage{30}
\doi{10.1214/11-STS382}

\makeatletter

\def\bsuffix #1{#1}

\makeatother

\begin{document}
\begin{frontmatter}
\vspace*{6pt}
\title{A Geometrical Explanation of~Stein~Shrinkage}
\runtitle{Stein Shrinkage}

\begin{aug}
\author[a]{\fnms{Lawrence D.} \snm{Brown}\corref{}\ead[label=e1]{lbrown@wharton.upenn.edu}}
\and
\author[b]{\fnms{Linda H.} \snm{Zhao}} 
\runauthor{L. D. Brown and L. H. Zhao}

\affiliation{University of Pennsylvania}

\address[a]{Lawrence D. Brown is Miers Busch Professor, Statistics Department,
The Wharton School, University of Pennsylvania, Philadelphia, PA
19010-6340, USA \printead{e1}.}
\address[b]{Linda H. Zhao is Professor, Statistics Department,
The Wharton School, University of Pennsylvania, Philadelphia, PA
19010-6340, USA.}

\end{aug}
\vspace*{-3pt}

%
\begin{abstract}
Shrinkage estimation has become a basic tool in the analysis of
high-dimensional data. Historically and conceptually a key development
toward this was the discovery of the inadmissibility of the usual
estimator of a multivariate normal mean.

This article develops a geometrical explanation for this
inadmissibility. By exploiting the spherical symmetry of the problem it
is possible to effectively conceptualize the multidimensional setting in
a two-dimensional framework that can be easily plotted and geometrically
analyzed. We begin with the heuristic explanation for inadmissibility
that was given by Stein [In \textit{Proceedings of the Third Berkeley
Sympo\-sium on Mathematical Statistics and Probability, 1954--1955, Vol. I}
(1956) 197--206, Univ. California Press]. Some geometric figures are
included to
make this reasoning more tangible. It is also explained why Stein's
argument falls short of yielding a proof of inadmissibility, even when
the dimension, $p$, is much larger than $p = 3$.

We then extend the geometric idea to yield increasingly persuasive
arguments for inadmissibility when $p \geq3$, albeit at the cost of
increased geometric and computational detail.
\end{abstract}

%

\end{frontmatter}

\section{Introduction}\label{sec1}

More than 50 years ago Stein (\citeyear{Ste56}) published his classic paper,
``Inadmissibility of the usual estimator for the mean of a multivariate
normal distribution.'' The title result is probably the most startling
statistical discovery of the past century. Erich Lehmann, who also
worked\vadjust{\goodbreak} on the admissibility question, more recently described how he was
``stunned with disbelief'' when Charles first told him of this result
(personal communication). Following the initial discovery James and
Stein (\citeyear{JamSte61}) presented their well-known shrinkage
estimator that
provides numerically significant improvement of risk relative to that of
the usual estimator.

[Hodges and Lehmann (\citeyear{HodLeh51}) and Girshick and Savage
(\citeyear{GirSav51}) had earlier
provided proofs of admissibility in the unidimensional problem;
Lehmann's student Blyth (\citeyear{Bly51}) had published another, more general,
argument for this same fact; and Lehmann and Stein (\citeyear
{LehSte53}) had produced a
proof of admissibility in a related one-dimensional hypothesis testing
setting.]

Stein (\citeyear{Ste56}) begins by describing the multivariate problem
and then
gives a heuristic, geometric argument intended to convince that the
usual estimator should be inadmissible if the dimension is sufficiently
large. The core of this argument will be repeated below,\vadjust{\goodbreak} with some
additional illustrations that hopefully help to clarify the situation.
The argument given by Stein provides insight into why inadmissibility
occurs in very high-dimensional problems. But it does not provide a
rationale for the fact that 3 is the critical dimension---admissibility
holds in dimension 1 and 2 but not in three or more dimensions. [Section
4 of Stein (\citeyear{Ste56}) contains an admissibility proof for two
dimensions.
See also Brown (\citeyear{Bro71}) and Brown and Fox (\citeyear{BroFox74}).]

$\!\!$The argument in the following note expands Stein's original heuristic
idea, clarifies the geometry, and provides justification for the fact
that 3 is the critical dimension. The argument is based on plane
geometry and some simple ``back-of-the-envelope'' Taylor series
expansions. As with Stein's argument, what is given here is not a proof.
It could undoubtedly be expanded into a proof, but without further
insight that proof would likely be similar to---and perhaps harder
than---the existing inadmissibility proofs in Stein (\citeyear{Ste56})
and Brown
(\citeyear{Bro66}). A slightly different geometrically based argument
is suggested
in Stein (\citeyear{St62}) and is additionally expanded in Brandwein and
Strawderman (\citeyear{Br90}). This argument is mentioned in Section \ref{sec3}.

Versions of this argument were presented in the 1960s in oral form
independently by L. Brown, by B. Efron, and perhaps by others. But so
far as we know the argument here does not appear in print. In addition,
we feel it is worthwhile to remind readers of the geometric rationale
underpinning Stein shrinkage in a form that displays that 3 is the
critical dimension.

\section{The Admissibility Problem}\label{sec2}

Let $\mathbf{X} = ( X_{1},\ldots,X_{p} )^{\prime}$ where $X_{i}$,
$i = 1,\ldots,p$, are independent normal variables with unknown means
$\theta_{1},\ldots,\theta_{p}$ and all with the same known variance,
$\sigma^{2}$. Without loss of generality, assume $\sigma^{2} = 1$. It is
desired to estimate $\bolds{\theta} = ( \theta_{1},\ldots,\theta_{p}
)^{\prime}$ with the quality of an estimate being measured through
squared error loss, $L( d,\theta) = \| d - \theta\|^{2} = \sum( d_{i} -
\theta_{i} )^{2}$. Let $\delta= \delta( \mathbf{X} )$ denote an
estimator. The risk function of $\delta$ is denoted by $R( \theta
;\delta) = E_{\theta} ( L( \delta( \mathbf{X} ) ) )$.

The ``usual'' estimator of $\bolds{\theta}$ is \textbf{X} itself, that
is, $\delta_{0}( \mathbf{X} ) = \mathbf{X}$. This estimator is intuitive
and has several appealing formal properties such as minimaxity,
best-invariance, maximum likelihood, etc. [See standard textbooks such
as Lehmann and Casella (\citeyear{LehCas98}) for discussion of these
properties.]

Prior to Stein (\citeyear{Ste56}) it had been firmly conjectured that
$\delta_{0}$
is admissible for any value of $p$. Admissibility means that there is no
other estimator that is better in the sense of risk---formally, that
there is no estimator $\delta'$ such that $R( \bolds{\theta} ;\delta'
) \le R( \bolds{\theta} ;\delta_{0} )$ with strict inequality at some
value of $\bolds{\theta}$. [Actually, though it is not important in
the sequel, we note that a well-known supplementary argument shows that
$\delta_{0}$ is inadmissible if and only if there is another estimator
that is always strictly better in the sense that $R( \bolds{\theta}
;\delta' ) < R( \bolds{\theta} ;\delta_{0} )$ for all $\bolds{\theta}.]$

What Stein proved in Sections 2--4 of Stein (\citeyear{Ste56})
is:

\begin{theorem*}[(Stein)] $\delta_{0}$ is admissible if and only if $p
\le2$.
\end{theorem*}

Our goal is to explain why $\delta_{0}$ is inadmissible when $p \ge3$.

\section{Spherical Symmetry}\label{sec3}

A spherically symmetric estimator is one that satisfies
%
\begin{equation}\label{eq1}
\delta( \mathbf{X} ) = \tau( \| \mathbf{X} \| )\mathbf{X}
\end{equation}
for some scalar function, $\tau$. Of course, $\delta_{0}$ is spherically
symmetric. We confine the search for alternatives to $\delta_{0}$ to the
collection of spherically symmetric estimators. Geometrically, these are
estimators that lie on the line through \textbf{X}, and whose distance
from the origin depends on $\| \mathbf{X} \|$. Such an estimator is
given as in (\ref{eq1})--(\ref{eq3}).

The restriction to spherically symmetric alternatives is intuitively
plausible. To support this intuition, Stein (\citeyear{Ste56}), Section~3, contains a
formal proof that $\delta_{0}$ is inadmissible if and only if there is a
spherically symmetric estimator which is better.

Once one has decided to restrict consideration only to spherically
symmetric estimators it is possible to correctly plot and study the
multivariate problem in a two- dimensional coordinate framework for the
sample space. One coordinate measures the sample in the direction of the
true parameter, $\theta$; the other coordinate is the length of the
orthogonal residual from this direction. This leads to the geometric
picture developed in the following section.

\section{Geometry for Spherically Symmetric Estimators}\label{sec4}

Only spherically symmetric estimators need to be considered. For such
estimators relevant distributions depend only on the magnitude of
$\bolds{\theta}$; the direction of the vector $\bolds{\theta}$ does
not matter. Formally, this means that after the constraint to
spherically symmetric estimators it suffices to consider the situation
when $\bolds{\theta}$ lies on the $\theta_{1}$-axis. So, assume
$\bolds{\theta} = ( \vartheta,0,\ldots,0 )^{\prime}$. Let $X = (
X_{1},X'_{( 2 )} )^{\prime}$ where $X_{( 2 )} \in\mathfrak{R}^{p - 1}$.
Geometrically, $X_{(2)}$ is the residual of \textbf{X} after projection
on the direction determined by $\bolds{\theta}$. Again, only the
length of $X_{( 2 )}$ matters, not its direction in the hyperplane
perpendicular to $\bolds{\theta}$. Hence, let $R = \| X_{( 2 )} \|$.
The relevant statistics for the observed sample can thus be rewritten as
%
\begin{eqnarray}\label{eq2}
&&\quad\mathbf{Z} = ( X_{1},R )\quad\mbox{with }X_{1} \sim N( \vartheta,1 )
,  R^{2} \sim\chi_{p - 1}^{2}
\nonumber
\\[-8pt]
\\[-8pt]
\nonumber
&&\quad\phantom{\mathbf{Z} = ( X_{1},R )\quad}\mbox{and }X_{1},R \mbox{ are independent.}
\end{eqnarray}
Spherically symmetric estimators as in (\ref{eq1}) are expressed similarly in
the $\mathbf{Z}$ coordinate system as
%
\begin{equation}\label{eq3}
\delta( \mathbf{Z} ) = \tau( \| \mathbf{Z} \| )\mathbf{Z}.
\end{equation}

The $\mathbf{Z}$ coordinate system is two-dimensional. Hen\-ce it can be
conveniently visualized geometrically. A~key feature of the
transformation leading from the original, \textbf{X}, system to the
$\mathbf{Z}$ system is that distances are preserved. In particular, for
spherically symmetric estimators
\[
\| \delta( \mathbf{X} ) - \theta\| = \| \delta( \mathbf{Z} ) - ( \vartheta,0 )
\|.
\]

Thus the squared error risks are the same in the two problems.

\begin{figure}

\includegraphics{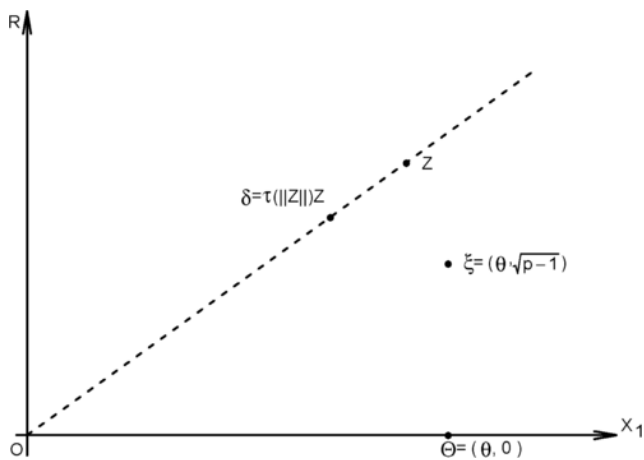}

  \caption{A typical observation in the $\mathbf{Z}=( X_{1},R )$ coordinate system.}\label{fig1}\vspace*{-3pt}
\end{figure}

Pictorially this can be plotted in standard planar coordinates, as
pictured in Figure \ref{fig1}. Figure \ref{fig1} shows a typical observation of \textbf{Z}
in the $( X_{1},R )$ coordinate system. It also represents a spherically
symmetric estimate corresponding to \textbf{Z}, as given by formulas~%
(\ref{eq1})--(\ref{eq3}). Pay special attention to the fact that this estimator is
on the line through \textbf{Z}. Figure \ref{fig1} also shows an additional
point\vadjust{\goodbreak}
$\xi= ( \xi_{1},\xi_{2} ) = ( \vartheta,\sqrt{p - 1} )$. This
represents the intuitive ``center'' of the distribution of \textbf{Z}.

In terms of Figure \ref{fig1} the statistical situation can be summarized as
follows: You observe $\mathbf{Z}$ with distribution as specified above.
You are constrained to use only spherically symmetric estimators that
lie on the line from the origin through $\mathbf{Z}$, as shown in the
plot. You want to find an estimator that is close to~$\Theta$ in terms
of squared distance. For the point shown on the plot it is fairly clear
that there are spherically symmetric estimates that are better than
just~$\mathbf{Z}$ alone. The point~$\delta$ shown on the plot is one such
better estimate. The goal of the remainder of the paper is to
substantiate that situations like that in the figure are \textit{on average}
sufficiently typical (at least when $p \geq3)$, and hence that
appropriate shrinkage estimators are better than $\mathbf{Z}$ itself.

[Note that $p - 1 = E( R^{2} )$. Hence it makes sense to think of\vadjust{\goodbreak}
$\sqrt{p - 1} = \xi_{2}$ as the center of the distribution of $R$. This
is not exactly either the mean or median of $R$, but it is sufficiently
close and is convenient for the following discussion. The exact mean of
$R$ is $E( R ) = \sqrt{2} \Gamma( p / 2 ) / \Gamma( ( p - 1 ) / 2 )$.
For $p = 5, 10, 17, 26$, respectively, this takes the values $E( R ) = 1.850,
2.918, 3.938, 4.950$ as compared to the values $\xi_{2} = \sqrt{p - 1} =
2, 3, 4, 5$. Asymptotically, $E( R ) = \sqrt{p - 1} - 1 / (4\sqrt{p - 1}
) + O( ( p - 1 )^{ - 3 / 2} ).]$

\begin{figure}

\includegraphics{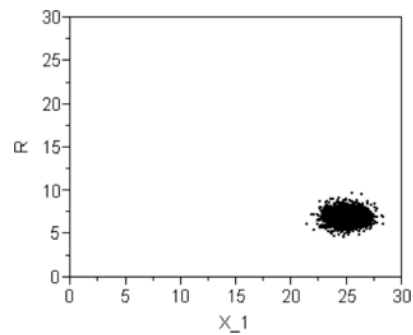}

  \caption{2000 observations of $\mathbf{Z}$ in the case $p = 20$ and
$\vartheta= 25$.}\label{fig2}
\end{figure}

Figure \ref{fig2} shows a typical sample of 2000 observations of \textbf{Z} in
the case $p = 20$ and $\vartheta= 25$. The dominant feature is that the
sample points are moderately tightly clustered about $\xi= (
25,\sqrt{19} )$ and hence are much closer to $\xi$ than they are to the
parameter point $\bolds{\theta} = ( \vartheta,0 )$.

\section{Stein's Heuristic Argument}\label{sec5}

It is fairly clear from pictures like Figure \ref{fig2} that shrinking the
observations somewhat toward the origin will often bring the estimator
closer to the true mean $\bolds{\theta} = ( \vartheta,0 )$. Even more
striking---consider what happens in a plot like Figure \ref{fig2} as $p
\to\infty$ for fixed $\bolds{\theta} = ( \vartheta,0 )$. Then the cloud
of points moves vertically upward. Eventually, virtually the entire
cloud lies outside the circle of radius $\| \bolds{\theta} \|$. To be
more precise
%
\begin{equation}\label{eq4}
\| X \|^{2} = \| \bolds{\theta} \|^{2} + p + O_{P}\bigl( \sqrt{p} \bigr)
\end{equation}
as $p \to\infty$ for any fixed $\bolds{\theta}$. This asymptotic fact
can be derived from the non-central chi-squared distribution of $\| X
\|^{2}$ or from a simple Taylor approximation as is done in Stein's
heuristic argument. Viewed another way, (\ref{eq4}) says that
%
\begin{eqnarray} \label{eq5}
\| \bolds{\theta} \| &=& \sqrt{\| X \|^{2} - p - O_{P}\bigl( \sqrt{p} \bigr)}
\nonumber
\\[-8pt]
\\[-8pt]
\nonumber
& =&
\| X \| - \frac{p + O_{P}( \sqrt{p} )}{2\| X \|}.
\end{eqnarray}

\begin{figure}

\includegraphics{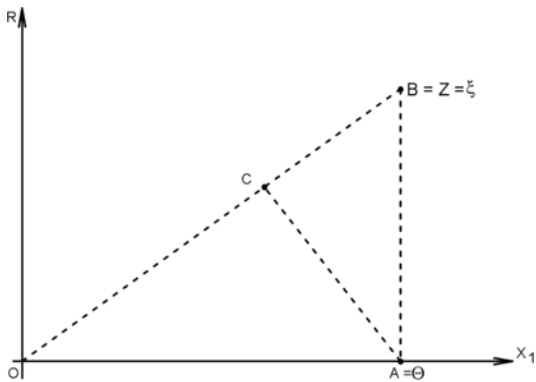}

  \caption{Geometry of the na\"{\i}ve optimal estimator: it shows the
origin, $\mathrm{O}$, the points $\mathrm{A} = ( \vartheta,0 )$, $\mathrm{B} = (
\vartheta,\sqrt{p - 1} )$ and $\mathrm{C}$, the projection of $\mathrm{A}$ on the line
$\overline{\mathrm{OB}}$.}\label{fig3}
\end{figure}

Any observation that lies outside of the sphere of radius $\|
\bolds{\theta} \|$ can be brought closer to $\bolds{\theta}$ by
shrinking it toward the origin so as to lie on the sphere. (Actually,
somewhat more shrinkage is desirable as will be clear from the
discussion of Figure \ref{fig3}, below.) This suggests that shrinkage by a factor
$\frac{p + O_{P}( \sqrt{p} )}{\| X \|}$ should be desirable as $p
\to\infty$. The argument in Stein (\citeyear{Ste56}) elaborates a
little further and
shows with a Taylor expansion that shrinkage by a factor $\frac{p + O( 1
)}{\| X \|}$ is still advantageous as $p \to\infty$ for any fixed
$\bolds{\theta}$. This motivates the use of the estimator $\delta_{p}(
X ) \buildrel\Delta\over= ( 1 - \frac{p}{\| X \|^{2}} )X$. This is
related to what is used in Stein (\citeyear{Ste56}) to prove
inadmissibility of the
usual estimator. The James--Stein (\citeyear{JamSte61}) estimator
$\delta_{p - 2}$ is
better than the usual one when $p \ge3$, as proved in that paper and
later in a more efficient manner through Stein's unbiased estimate of
risk in Stein (\citeyear{Ste74,Ste81}).

[Since $p \to\infty$ the difference between the factor $p / \| X \|^{2}$
in this argument and the factor $( p - 2 ) / \| X \|^{2}$ in James and
Stein (\citeyear{JamSte61}) is irrelevant. For fixed $p$ it can be
shown by the
arguments mentioned above that $\delta_{p}$ dominates $\delta_{0}$
whenever $p \ge4$.]

Stein (\citeyear{Ste56}) writes that ``With some additional precision this
[heuristic argument] could be made $\ldots[$in]to$\ldots$ a proof that
for sufficiently large [$p]$ the usual estimator is inadmissible.'' This
is the type of exaggeration that may be excused by the above being only
meant as a heuristic argument. In fact much more than ``some additional
precision'' is needed to prove the usual estimator is inadmissible for
sufficiently large $p$. The reason that the above does not easily yield
a proof of inadmissibility is that it only holds for any fixed
$\bolds{\theta}$ as $p \to\infty$. It does not hold uniformly in
$\bolds{\theta}$, but a uniform argument is needed in order to prove
inadmissibility.

To be more precise, for any $p$ no matter how large,
$\inf_{\bolds{\theta}} \{ P_{\bolds{\theta}} ( \| X \| \ge\|
\bolds{\theta} \| ) \} = 1 / 2$, rather than approaching~1 as is
implicitly suggested within the heuristic argument, and as would be
needed to easily convert the heuristic argument into a proof.

Hence a more elaborate argument is needed to prove that the usual
estimator is inadmissible. The following discussion presents a heuristic
argument for inadmissibility that is consistent with the geometric
insight in Stein's motivation.

\section{Desired Amount of Shrinkage; Typical Observation}\label{sec6}

Figures \ref{fig1} and \ref{fig2} show that the observations are close to $\xi= (
\vartheta,\sqrt{p - 1} )$, whereas the estimate should be as close as
possible to $\bolds{\theta} = ( \vartheta,0 )$. Figure \ref{fig3} illustrates
the geometry of this situation when $\mathbf{Z} = ( \vartheta,\sqrt{p -
1} )$. It shows the origin (O), the point $\mathrm{A} = \bolds{\theta}
= ( \vartheta,0 )$ which is the desired target of the estimate, and the
point $\mathrm{B} = \xi= ( \vartheta,\sqrt{p - 1} )$ which is a typical
observation. For such an observation any spherically symmetric estimator
must be on the line $\overline{\mathrm{OB}}$. The point C in Figure~\ref{fig3} is
the point on that line which is closest to the desired target, A. A
similar triangles yield that
\[
\frac{| \overline{\mathrm{AB}} |}{|
\overline{\mathrm{OB}} |} = \frac{| \overline{\mathrm{BC}}
|}{| \overline{\mathrm{AB}} | },
\]
where $| \overline{\mathrm{AB}} |$ denotes the length of the segment
$\overline{\mathrm{AB}}$, etc. Simplifying yields
%
\begin{equation}\label{eq6}
| \overline{\mathrm{BC}} | = \frac{| \overline{\mathrm{AB}} |^{2}}{|
\overline{\mathrm{OB}} |} = \frac{p - 1}{\| \xi\|}
.
\end{equation}

The point C is the best estimate based on an observation at $\mathrm{B}
= \xi$. By (\ref{eq6}) it can be written as
\[
\mathrm{B} = \biggl( 1 - \frac{p - 1}{\| \xi\|^{2}}
\biggr)\xi.
\]

By comparison with (\ref{eq1}), this suggests that the optimal spherically
symmetric estimator will be the Na\"{\i}ve Geometrically Optimal
estimator
%
\begin{equation}\label{eq7}
\delta_{\mathrm{NGO}}( \mathbf{Z} ) = \biggl( 1 - \frac{p - 1}{\|
\mathbf{Z} \|^{2}} \biggr)\mathbf{Z}.
\end{equation}

The discussion leading to (\ref{eq7}) suggests that
\[
\delta_{p - 1}( X ) = \biggl( 1 - \frac{p - 1}{\| X
\|^{2}} \biggr)X
\]
should dominate $\delta_{0}$. The above motivation and construction of
$\delta_{\mathrm{NGO}}$ does not suffer from the defect noted above in
Stein's original heuristics---it does not require $p \to\infty$ for each
fixed $\vartheta$. However, it suggests that $\delta_{0}$ is
inadmissible even for $p = 2$. This suggestion is not correct; and so a
more careful heuristic argument is needed to get a better description of
the relevant geometry.

\section{Stochastic Variation}\label{sec7}

\begin{figure}

\includegraphics{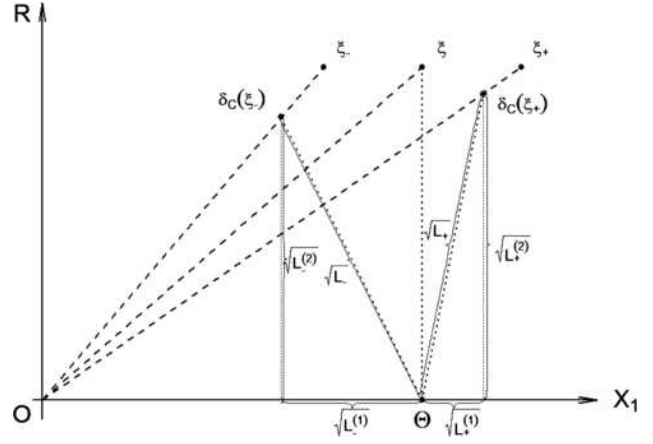}

  \caption{The values of $\xi_{ \pm}$ and their respective estimates.}\label{fig4}
\end{figure}

The estimator $\delta_{\mathrm{NGO}}$ in (\ref{eq7}) is only optimal at $\xi=
( \vartheta,\sqrt{p - 1} )$, the central point of the distribution of~\textbf{Z}.
Of course,~\textbf{Z} is not identically $\xi$, but is only
stochastically close to $\xi$. The calculation leading to (\ref{eq7}) is only
approximate, not exact. There is a small price in accuracy to be paid in
order to accommodate the stochastic variation of \textbf{Z}. In order to
better understand the composition of this price consider a particular
pair of equally likely possible points for \textbf{Z}. These points are
labeled $\xi_{\mathbf{ +}} , \xi_{ -}$ in Figure \ref{fig4}. They are defined as
\[
\xi_{ \pm} = \bigl( \vartheta\pm1, \sqrt{p - 1} \bigr).
\]

These points exhibit typical stochastic variation in the direction of
$\theta= ( \vartheta,0 )$ since their mean and mean squared distance in
that direction match those of the full distribution. While they do not
accurately model the stochastic variation in the direction orthogonal to
$\theta= ( \vartheta,0 )$, it turns out that this additional
variability is only of secondary importance. Thus, we will ignore the
effect of this orthogonal variation for now. It becomes clear from the
exact expression discussed later at (\ref{eq14})--(\ref{eq15}) that the orthogonal
variation is indeed of secondary importance in calculation of the
difference in risks.

Note that $L_{ +} < \| \xi_{ +} - \vartheta\|^{2} = \| \xi_{ -} -
\vartheta\|^{2}$ but $L_{ -}$ can be $> \| \xi_{ -} - \vartheta\|^{2}$.
Calculations in the test show that $\frac{1}{2}(L_{ +} + L_{ -} ) <
\Vert \xi_{ \pm} - \vartheta\Vert ^{2}$when $0 < {C} < 2( {p} - 2
)$.\vadjust{\goodbreak}

In order to allow for additional discussion consider the general form
%
\begin{equation}\label{eq8}
\delta_{C}( \mathbf{Z} ) = \biggl( 1 - \frac{C}{\| \mathbf{Z} \|^{2}} \biggr)\mathbf{Z}.
\end{equation}

The case $C = p - 1$ is motivated by the preceding geometric argument.
But the following calculations suggest that because of the stochastic
variation modeled through $\xi_{{ +}}$, $\xi_{ -}$ a preferable
choice is $C = p - 2$, as in the ordinary James--Stein
estimator.

Break down the risk into two components corresponding to the directions
determined by the coordinates $\mathbf{Z}=( X_{1},R )$. This is similar to the
suggestion in Stein (\citeyear{Ste56}), remark (vii). Related
calculations
are described in Efron and Morris (\citeyear{EfrMor71}). Let $L_{ \pm}$
denote the
squared error from an observation at one of the two equally likely
points $\xi_{{ +}}$, $\xi_{ -}$ , respectively,
\begin{eqnarray*}
L_{ \pm} &=& \biggl[\biggl ( 1 - \frac{C}{\| \xi_{ \pm} \|^{2}} \biggr)(
\vartheta\pm1 ) - \vartheta\biggr]^{2} \\[1pt]
&&{}+ \biggl[ \biggl( 1 - \frac{C}{\| \xi_{ \pm}
\|^{2}} \biggr)\sqrt{p - 1} - 0 \biggr]^{2} \\[1pt]
& =& \biggl[ \pm1 - \frac{C}{\| \xi_{ \pm}
\|^{2}}( \vartheta\pm1 ) \biggr]^{2} \\[1pt]
&&{}+ \biggl( 1 - \frac{C}{\| \xi_{ \pm} \|^{2}}
\biggr)^{2}( p - 1 ) \\[1pt]
&\buildrel\Delta\over=& L_{ \pm} ^{(1)} + L_{ \pm}
^{(2)},\quad \mbox{say}.
\end{eqnarray*}\looseness=0

Let $R_{| \xi_{ \pm}}( \bolds{\theta} ,\delta_{C} )$ denote the
conditional risk given that $Z = \xi_{{ +}}$ or $\xi_{
-}$. Then
\begin{eqnarray*}
R_{| \xi_{ \pm} } &=& \tfrac{1}{2}\bigl( L_{ +} ^{(1)} + L_{
-} ^{(1)} \bigr) + \tfrac{1}{2}\bigl( L_{ +} ^{(2)} + L_{ -} ^{(2)}
\bigr) \\
&\buildrel\Delta\over=& R_{| \xi_{ \pm} .}^{(1)} +
R_{| \xi_{ \pm} .}^{(2)}, \quad\mbox{say}.\vadjust{\goodbreak}
\end{eqnarray*}

Is $\delta_{C}$ better than $\delta_{0}$ for this conditional problem?
To examine this we look at the coordinate-wise difference in conditional
risks. For $\delta_{0}$ the coordinate-wise risks are 1 and $p - 1$,
respectively. Hence the coordinate-wise differences are
\begin{eqnarray*}
&&1 - R_{| \xi_{ \pm} }^{(1)} = \frac{1}{2}\biggl(
\frac{2C( \vartheta+ 1 )}{\| \xi_{ +} \|^{2}} -
\frac{C^{2}( \vartheta+ 1 )^{2}}{\| \xi_{ +}
\|^{4}} \\
&&\phantom{1 - R_{| \xi_{ \pm} }^{(1)} = \frac{1}{2}\biggl(}{}- \frac{2C( \vartheta- 1 )}{\| \xi_{ -}
\|^{2}} - \frac{C^{2}( \vartheta- 1 )^{2}}{\| \xi
_{ -} \|^{4}} \biggr)
\end{eqnarray*}
and
%
\begin{eqnarray*}
&&( p - 1 ) - R_{| \xi_{ \pm} }^{(2)} \\
&&\quad= \frac{1}{2}( p - 1 )\biggl( \biggl(
\frac{2C}{\| \xi_{ +} \|^{2}} + \frac{2C}{\| \xi_{ -} \|^{2}} \biggr)\\
&&\hspace*{46pt}\qquad{} - \biggl(
\frac{C^{2}}{\| \xi_{ +} \|^{4}} + \frac{C^{2}}{\| \xi_{ -} \|^{4}} \biggr)
\biggr).
\end{eqnarray*}

In order to better interpret this expression rearrange terms so as to
write the improvement of $\delta_{C}$ over $\delta_{0}$ in this
conditional problem as
%
\begin{eqnarray}\label{eq9}
 \Delta_{| \xi_{ \pm} }& \buildrel\Delta\over=& p -
R_{| \xi_{ \pm} } \nonumber\\
& =& C\vartheta\biggl( \frac{1}{\| \xi_{ +} \|^{2}} -
\frac{1}{\| \xi_{ -} \|^{2}} \biggr)\nonumber\\
&&{} + Cp\biggl( \frac{1}{\| \xi_{ +} \|^{2}} +
\frac{1}{\| \xi_{ -} \|^{2}} \biggr) \nonumber\\
&&{}- \frac{1}{2}C^{2}\biggl( \frac{( \vartheta+
1 )^{2} + p - 1}{\| \xi_{ +} \|^{4}} \\
&&\hspace*{40pt}{}+ \frac{( \vartheta- 1 )^{2} + p
- 1}{\| \xi_{ -} \|^{4}} \biggr)\nonumber \\
&=& C\vartheta\biggl( \frac{1}{\| \xi_{ +}
\|^{2}} - \frac{1}{\| \xi_{ -} \|^{2}} \biggr)\nonumber\\
&&{} + \biggl( Cp - \frac{1}{2}C^{2} \biggr)\biggl(
\frac{1}{\| \xi_{ +} \|^{2}} + \frac{1}{\| \xi_{ -} \|^{2}}
\biggr)\nonumber
\end{eqnarray}
since $\xi_{ \pm} = ( \vartheta\pm1 )^{2} + p - 1$.

\textit{If it were so} that $\| \xi_{ +} \|^{2} = \| \xi_{ -} \|^{2}$
\textit{then} the first major term on the right of (\ref{eq9}) would be${}=$0, and the difference in (\ref{eq9}) would be positive for any $0 < C
< 2p$. In particular, for any $p \ge2$ it \textit{would be} positive
for $C = p - 1$. (It could even be positive for $p = 1!)$ This of
course makes no sense as a statistical solution and only confirms that
it provides an incorrect insight to ignore that $\| \xi_{ +} \|^{2} >
\| \xi_{ -} \|^{2}$.

Now, look at (\ref{eq9}), and take into account that\break $\| \xi_{ +}
\|^{2}
> \| \xi_{ -} \|^{2}$. Then, $\frac{1}{\| \xi_{ +} \|^{2}} -
\frac{1}{\| \xi _{ -} \|^{2}} < 0$, and the first term\vadjust{\goodbreak} on the right of
(3.9) is negative and partially compensates for the remaining term
which is positive when \mbox{$C = p - 1$}. In more detail,
\begin{eqnarray*}
\frac{1}{\| \xi_{ +} \|^{2}} - \frac{1}{\| \xi_{ -} \|^{2}} &=&
\frac{\| \xi_{ -} \|^{2} - \| \xi_{ +} \|^{2}}{\| \xi_{ +} \|^{2}\|
\xi_{ -}
\|^{2}}\\
& = &- 4\frac{\vartheta} {\| \xi_{ +} \|^{2}\| \xi_{ -} \|^{2}}, \\
\frac{1}{\| \xi_{ +} \|^{2}} + \frac{1}{\| \xi_{ -} \|^{2}}& = &\frac{\|
\xi_{ -} \|^{2} + \| \xi_{ +} \|^{2}}{\| \xi_{ +} \|^{2}\| \xi_{ -}
\|^{2}} = 2\frac{\vartheta^{2} + p}{\| \xi_{ +} \|^{2}\| \xi_{ -}
\|^{2}} .
\end{eqnarray*}

Hence the difference in conditional risks for $p \ge2$ is
%
\begin{eqnarray}\label{eq10}
\Delta_{| \xi_{ \pm} }& =& p - R_{| \xi_{ \pm} }\nonumber \\
&= &\frac{2}{\| \xi_{ +} \|^{2}\| \xi_{ -} \|^{2}}
\nonumber
\\[-8pt]
\\[-8pt]
\nonumber
&&{}\cdot\biggl( \biggl( C( p
- 2 )
- \frac{C^{2}}{2} \biggr)\vartheta^{2} + \biggl( Cp - \frac{C^{2}}{2} \biggr)p \biggr) \\
&> &\frac{2( \vartheta^{2} + p )}{\| \xi_{ +} \|^{2}\| \xi_{ -}
\|^{2}}\biggl( C( p - 2 ) - \frac{C^{2}}{2} \biggr) .\nonumber
\end{eqnarray}

It follows that the difference in conditional risks is positive so long
as $0 < C < 2( p - 2 )$. In particular the difference is positive for
$p \ge3$ and $C = p - 1$, the value motivated by the geometric argument
centered on Figure \ref{fig3}. On the other hand, the best choice of
constant in (\ref{eq10}) is the slightly smaller value $C = p - 2$. The
improvement in risks is not as great as that suggested in the argument
around Figure \ref{fig3}, and this can be considered as a necessary
penalty due to the randomness in~\textbf{X}. In summary, the result in
(\ref{eq10}) provides a heuristic motivation for inadmissibility to
hold whenever$p \ge3$.

\section{What Can Be Proved}\label{sec8}

Note in (\ref{eq10}) that the three terms in the leading fraction are all
approximately equal; that is, $\vartheta^{2} + p \approx\| \xi_{ +}
\|^{2} \approx\| \xi_{ +} \|^{2}$. Hence the argument leading to (\ref{eq10})
suggests that the unconditional difference in risks, $\Delta= R(
\bolds{\theta} ,\delta_{0} ) - R( \bolds{\theta} ,\delta_{C} )$, will
be well approximated as
%
\begin{eqnarray}\label{eq11}
\Delta&=& R( \bolds{\theta} ,\delta_{0} ) - R( \bolds{\theta}
,\delta_{C} )
\nonumber
\\[-8pt]
\\[-8pt]
\nonumber
& \approx&\frac{2}{\| \bolds{\theta} \|^{2} + p}\biggl( C( p - 2 )
- \frac{C^{2}}{2} \biggr).
\end{eqnarray}
The quality of this approximation improves as $\| \bolds{\theta} \|
\to\infty$ in the sense that
%
\begin{eqnarray}\label{eq12}
\Delta\sim\frac{2}{\| \bolds{\theta} \|^{2} + p}\biggl( C( p - 2 ) -
\frac{C^{2}}{2} \biggr)
\nonumber
\\[-8pt]
\\[-8pt]
\eqntext{\mbox{as }\| \bolds{\theta} \| \to\infty.}
\end{eqnarray}

The preceding arguments can be refined to prove the assertion in (\ref{eq12}).
This is essentially the path followed by Stein in his original argument
in Stein (\citeyear{Ste56}). In order to allow calculations accurate
only for large
$\| \bolds{\theta} \|$, Stein replaced $\delta_{C}$ with the estimator
\[
\delta_{C;a} = \biggl( 1 - \frac{C}{a + \| \mathbf{X} \|^{2}} \biggr)\mathbf{X}.
\]

Then an exact Taylor expansion that can be conside\-red as an elaboration
of the above calculations yields
%
\begin{eqnarray}\label{eq13}
&&R( \bolds{\theta} ,\delta_{0} ) - R( \bolds{\theta} ,\delta_{C;a}
)\nonumber \\[-2pt]
&&\quad= \frac{2}{a + \| \bolds{\theta} \|^{2}}\biggl( C( p - 2 ) -
\frac{C^{2}}{2} \biggr)\\[-2pt]
&&\qquad{} + o\biggl( \frac{1}{a + \| \bolds{\theta} \|^{2}}
\biggr)\nonumber
\end{eqnarray}
uniformly in $\| \bolds{\theta} \|$. It follows that $\delta_{0}$ is
inadmissible.

The argument in Stein (\citeyear{Ste56}) for (\ref{eq13}) involves only
low-order moments
of $\mathbf{X}-\bolds{\theta}$. Hence it can be generalized from the normal distribution
setting to apply to more general location parameter problems. It can
also be adapted to apply (with modifications) to problems in which the
loss function is not squared error. Such generalizations appear in Brown
(\citeyear{Bro66}).

When one considers only the normal distribution setting, then
%
\begin{eqnarray}\label{eq14}
\Delta&=& R( \bolds{\theta} ,\delta_{0} ) - R( \bolds{\theta}
,\delta_{C} )
\nonumber
\\[-10pt]
\\[-10pt]
\nonumber
&=& E_{\theta} \biggl( \frac{1}{\| \mathbf{X} \|^{2}} \biggr)\biggl( C( p - 2
) - \frac{C^{2}}{2} \biggr).
\end{eqnarray}

This result is proved but not explicitly stated in James and Stein
(\citeyear{JamSte61}). It is explicitly stated and proved using the
unbiased estimate
of the risk in Stein (\citeyear{Ste74,Ste81}).

Note that
%
\begin{equation}\label{eq15}
E_{\bolds{\theta}} \biggl( \frac{1}{\| \mathbf{X} \|^{2}} \biggr)
\approx\frac{1}{E_{\bolds{\theta}} ( \| \mathbf{X} \|^{2} )} =
\frac{1}{\| \bolds{\theta} \|^{2} + p},
\end{equation}
with the approximation being quite close except\break when $\| \bolds{\theta}
\|$ is small. Hence the heuristic approximation in~(\ref{eq10}) and (\ref{eq11}) is
quite close to the truth. This validates the heuristic idea to
approximate the unconditional difference in risks by the conditional
difference given $\bolds{\theta} = \xi_{\mathbf{ +}}$, $\xi_{ -}$ .

\section*{Acknowledgment}
Research supported in part by NSF Grant DMS-10-07657.

%


\begin{thebibliography}{14}

\bibitem[\protect\citeauthoryear{}{1951}]{Bly51}
%
\begin{barticle}[mr]
\bauthor{\bsnm{Blyth},~\bfnm{Colin~R.}\binits{C.~R.}}
(\byear{1951}).
\btitle{On minimax statistical decision procedures and their admissibility}.
\bjournal{Ann. Math. Statist.}
\bvolume{22}
\bpages{22--42}.
\bid{issn={0003-4851}, mr={0039966}}
\bptok{imsref}%
\end{barticle}
%
\endbibitem

\bibitem[\protect\citeauthoryear{}{1990}]{Br90}
%
\begin{barticle}[mr]
\bauthor{\bsnm{Brandwein},~\bfnm{A.~C.}\binits{A.~C.}} \AND
\bauthor{\bsnm{Strawderman},~\bfnm{W.~E.}\binits{W.~E.}}
(\byear{1990}).
\btitle{Stein estimation: The spherically symmetric case}.
\bjournal{Statist. Sci.}
\bvolume{5}
\bpages{356--369}.
\bptok{imsref}%
\end{barticle}
%
\endbibitem


\bibitem[\protect\citeauthoryear{}{1966}]{Bro66}
%
\begin{barticle}[mr]
\bauthor{\bsnm{Brown},~\bfnm{Lawrence~David}\binits{L.~D.}}
(\byear{1966}).
\btitle{On the admissibility of invariant estimators of one or more location
parameters}.
\bjournal{Ann. Math. Statist.}
\bvolume{37}
\bpages{1087--1136}.
\bid{issn={0003-4851}, mr={0216647}}
\bptok{imsref}%
\end{barticle}
%
\endbibitem

\bibitem[\protect\citeauthoryear{}{1971}]{Bro71}
%
\begin{barticle}[mr]
\bauthor{\bsnm{Brown},~\bfnm{L.~D.}\binits{L.~D.}}
(\byear{1971}).
\btitle{Admissible estimators, recurrent diffusions, and insoluble boundary
value problems}.
\bjournal{Ann. Math. Statist.}
\bvolume{42}
\bpages{855--903}.
\bid{issn={0003-4851}, mr={0286209}}
[\bnote{Correction. \textit{Ann. Statist.}
\textbf{1} (1973) 594--596.}%
\bid{mr={0362592}}]
\bptok{imsref}%
\end{barticle}
%
\endbibitem

\bibitem[\protect\citeauthoryear{}{1974}]{BroFox74}
%
\begin{barticle}[mr]
\bauthor{\bsnm{Brown},~\bfnm{Lawrence~D.}\binits{L.~D.}} \AND
\bauthor{\bsnm{Fox},~\bfnm{Martin}\binits{M.}}
(\byear{1974}).
\btitle{Admissibility in statistical problems involving a location or scale
parameter}.
\bjournal{Ann. Statist.}
\bvolume{2}
\bpages{807--814}.
\bid{issn={0090-5364}, mr={0370850}}
\bptok{imsref}%
\end{barticle}
%
\endbibitem

\bibitem[\protect\citeauthoryear{}{1971}]{EfrMor71}
%
\begin{barticle}[mr]
\bauthor{\bsnm{Efron},~\bfnm{Bradley}\binits{B.}} \AND
\bauthor{\bsnm{Morris},~\bfnm{Carl}\binits{C.}}
(\byear{1971}).
\btitle{Limiting the risk of {B}ayes and empirical {B}ayes estimators. {I}.
{T}he {B}ayes case}.
\bjournal{J. Amer. Statist. Assoc.}
\bvolume{66}
\bpages{807--815}.
\bid{issn={0162-1459}, mr={0323014}}
\bptok{imsref}%
\end{barticle}
%
\endbibitem

\bibitem[\protect\citeauthoryear{}{1951}]{GirSav51}
%
\begin{binproceedings}[mr]
\bauthor{\bsnm{Girshick},~\bfnm{M.~A.}\binits{M.~A.}} \AND
\bauthor{\bsnm{Savage},~\bfnm{L.~J.}\binits{L.~J.}}
(\byear{1951}).
\btitle{Bayes and minimax estimates for quadratic loss functions}.
In \bbooktitle{Proceedings of the {S}econd {B}erkeley {S}ymposium on
{M}athematical {S}tatistics and {P}robability, 1950}
\bpages{53--73}.
\bpublisher{Univ. California Press}, \baddress{Berkeley}.
\bid{mr={0045365}}
\bptok{imsref}%
\end{binproceedings}
%
\endbibitem

\bibitem[\protect\citeauthoryear{}{1951}]{HodLeh51}
%
\begin{binproceedings}[mr]
\bauthor{\bsnm{Hodges},~\bfnm{J.~L.}\binits{J.~L.}, \bsuffix{Jr.}} \AND
\bauthor{\bsnm{Lehmann},~\bfnm{E.~L.}\binits{E.~L.}}
(\byear{1951}).
\btitle{Some applications of the {C}ram\'er-{R}ao inequality}.
In \bbooktitle{Proceedings of the {S}econd {B}erkeley {S}ymposium on
{M}athematical {S}tatistics and {P}robability, 1950}
\bpages{13--22}.
\bpublisher{Univ. California Press}, \baddress{Berkeley}.
\bid{mr={0044795}}
\bptok{imsref}%
\end{binproceedings}
%
\endbibitem

\bibitem[\protect\citeauthoryear{}{1961}]{JamSte61}
%
\begin{bincollection}[mr]
\bauthor{\bsnm{James},~\bfnm{W.}\binits{W.}} \AND
\bauthor{\bsnm{Stein},~\bfnm{Charles}\binits{C.}}
(\byear{1961}).
\btitle{Estimation with quadratic loss}.
In \bbooktitle{Proc. 4th {B}erkeley {S}ympos. {M}ath. {S}tatist. and {P}rob.,
{V}ol. {I}}
\bpages{361--379}.
\bpublisher{Univ. California Press}, \baddress{Berkeley, CA.}
\bid{mr={0133191}}
\bptnote{check year}%
\bptok{imsref}%
\end{bincollection}
%
\endbibitem

\bibitem[\protect\citeauthoryear{}{1998}]{LehCas98}
%
\begin{bbook}[mr]
\bauthor{\bsnm{Lehmann},~\bfnm{E.~L.}\binits{E.~L.}} \AND
\bauthor{\bsnm{Casella},~\bfnm{George}\binits{G.}}
(\byear{1998}).
\btitle{Theory of Point Estimation},
\bedition{2nd} ed.
\bpublisher{Springer}, \baddress{New York}.
\bid{mr={1639875}}
\bptok{imsref}%
\end{bbook}
%
\endbibitem

\bibitem[\protect\citeauthoryear{}{1953}]{LehSte53}
%
\begin{barticle}[mr]
\bauthor{\bsnm{Lehmann},~\bfnm{E.~L.}\binits{E.~L.}} \AND
\bauthor{\bsnm{Stein},~\bfnm{C.~M.}\binits{C.~M.}}
(\byear{1953}).
\btitle{The admissibility of certain invariant statistical tests
involving a
translation parameter}.
\bjournal{Ann. Math. Statist.}
\bvolume{24}
\bpages{473--479}.\break
\bid{issn={0003-4851}, mr={0056249}}
\bptok{imsref}%
\end{barticle}
%
\endbibitem

\bibitem[\protect\citeauthoryear{}{1956}]{Ste56}
%
\begin{binproceedings}[mr]
\bauthor{\bsnm{Stein},~\bfnm{Charles}\binits{C.}}
(\byear{1956}).
\btitle{Inadmissibility of the usual estimator for the mean of a multivariate
normal distribution}.
In \bbooktitle{Proceedings of the {T}hird {B}erkeley {S}ymposium on
{M}athematical {S}tatistics and {P}robability, 1954--1955, Vol. {I}}
\bpages{197--206}.
\bpublisher{Univ. California Press}, \baddress{Berkeley}.
\bid{mr={0084922}}
\bptok{imsref}%
\end{binproceedings}
%
\endbibitem




\bibitem[\protect\citeauthoryear{}{1973}]{Ste74}
%
\begin{binproceedings}[mr]
\bauthor{\bsnm{Stein},~\bfnm{Charles}\binits{C.}}
(\byear{1973}).
\btitle{Estimation of the mean of a multivariate normal distribution}.
In \bbooktitle{Proceedings of the {P}rague {S}ymposium on {A}symptotic
{S}tatistics ({C}harles {U}niv., {P}rague, 1973), {V}ol. {II}}
\bpages{345--381}.
\bpublisher{Charles Univ.}, \baddress{Prague}.
\bid{mr={0381062}}
\bptnote{check year}%
\bptok{imsref}%
\end{binproceedings}
%
\endbibitem



\bibitem[\protect\citeauthoryear{}{1962}]{St62}
%
\begin{barticle}[mr]
\bauthor{\bsnm{Stein},~\bfnm{Charles~M.}\binits{C.~M.}}
(\byear{1962}).
\btitle{Confidence sets for the mean of a multivariate normal distribution}.
\bjournal{J. R. Statist. Soc. Ser. B Stat. Methodol.}
\bvolume{24}
\bpages{265--296}.
\bptok{imsref}%
\end{barticle}
%
\endbibitem


\bibitem[\protect\citeauthoryear{}{1981}]{Ste81}
%
\begin{barticle}[mr]
\bauthor{\bsnm{Stein},~\bfnm{Charles~M.}\binits{C.~M.}}
(\byear{1981}).
\btitle{Estimation of the mean of a multivariate normal distribution}.
\bjournal{Ann. Statist.}
\bvolume{9}
\bpages{1135--1151}.
\bid{issn={0090-5364}, mr={0630098}}
\bptok{imsref}%
\end{barticle}
%
\endbibitem

\end{thebibliography}
\end{document}